\documentclass[manuscript]{aastex}

\usepackage{color}
\usepackage{graphicx}
\usepackage{graphicx,amsmath}

\shorttitle{Origin of the chaotic motion of Atlas}
\shortauthors{Renner et al.}

\begin{document}

\title{Origin of the chaotic motion of the Saturnian satellite Atlas}

\author{S. Renner\altaffilmark{1,2}, 
N.J. Cooper\altaffilmark{3}, 
M. El Moutamid\altaffilmark{4}, 
B. Sicardy\altaffilmark{5},
A. Vienne\altaffilmark{1,2}, 
C.D. Murray\altaffilmark{3}, 
M. Saillenfest\altaffilmark{2,6}
}

\altaffiltext{1}{Universit\'e Lille 1, Laboratoire d'Astronomie de Lille (LAL), 
1 impasse de l'Observatoire, 59000 Lille, France}
\altaffiltext{2}{Institut de M\'ecanique C\'eleste et de Calcul des Eph\'em\'erides (IMCCE), 
Observatoire de Paris, CNRS UMR 8028, 77 avenue Denfert-Rochereau, 75014 Paris, France}
\altaffiltext{3}{Astronomy Unit, School of Physics and Astronomy, Queen Mary University of London, 
Mile End Road, London, E1 4NS, UK}
\altaffiltext{4}{Department of Astronomy, Cornell University, Ithaca, NY 14853, USA}
\altaffiltext{5}{LESIA/Observatoire de Paris, PSL, CNRS UMR 8109, 
University Pierre et Marie Curie, University Paris-Diderot, 
5 place Jules Janssen, 92195 Meudon C\'edex, France}
\altaffiltext{6}{Department of Mathematics, University of Pisa, Italy}

\catcode`±=\active
\def±{\phantom{$-$}}
\catcode`?=\active
\def?{\phantom{0}}

\begin{abstract}
     We revisit the dynamics of Atlas. Using $Cassini$ ISS astrometric observations spanning 
	  February 2004 to August 2013, \citet{Coop15} found evidence that Atlas is currently 
	  perturbed by both 
	  a 54:53 corotation eccentricity resonance (CER) and a 54:53 Lindblad eccentricity resonance (LER) 
	  with Prometheus. 
      They demonstrated that the orbit of Atlas is chaotic, with a Lyapunov time of order 10 years,  
	  as a direct consequence of the coupled resonant interaction (CER/LER) 
	  with Prometheus. 
	  Here we investigate the interactions between the two resonances using the CoraLin
      analytical model 
	  \citep{Elmout14}, showing that the chaotic zone fills almost all the corotation sites occupied by 
	  the satellite's orbit. Four 70:67 apse-type mean motion resonances with Pandora are also 
	  overlapping, but these resonances have a much weaker effect. Frequency analysis allows 
	  us to highlight the coupling between the 54:53 resonances, and confirms that a simplified 
	  system including the perturbations due to Prometheus and Saturn's oblateness only 
	  captures the essential features of the dynamics.

\end{abstract}


\clearpage

\section{Introduction}

Thanks to the success of observing campaigns such as 
those using the Imaging Science Subsystem (ISS) of the $Cassini$ orbiter, 
the short-term dynamical evolution of the small inner Saturnian satellites 
can now be studied with a high accuracy. This is important in order to give
constraints on the physical origin and orbital evolution of these moons. 

This work focuses on Atlas, the closest satellite to the outer edge of Saturn's A ring.
This satellite lies in a complex dynamical environment involving various mean motion resonances : 
		\begin{itemize}
            \item $54:53$ resonances between Atlas and Prometheus \citep{Spit06,Coop15}
			\item $70:67$ resonances between Atlas and Pandora \citep{Spit06}
			\item $121:118$ resonances between Prometheus and Pandora, leading to chaotic interactions
			every 6.2 years at closest approach (apse anti-alignment)
			    \citep{Gold03a,Gold03b,Ren03,Coop04,Ren05}
				\item $1:1$ co-orbital resonance between Janus and Epimetheus
            \item $17:15$ or $21:19$ resonances between Prometheus or Pandora and Epimetheus \citep{Coop04}
			\item $3:2$ near-resonances between Pandora and Mimas \citep{French03}
		\end{itemize}
In fact, we show in this paper that Atlas is mainly perturbed by the $54:53$ resonant perturbations
from Prometheus. The latter involve two resonance arguments :
$\Psi_C=$54$\lambda_{PR}$$-$53$\lambda_{AT}$$-$$\varpi_{PR}$ and 
$\Psi_L=$54$\lambda_{PR}$$-$53$\lambda_{AT}$$-$$\varpi_{AT}$, 
corresponding to the Corotation Eccentricity Resonance (hereafter, CER) and the 
Lindblad Eccentricity Resonance (LER), respectively (see \citet{Elmout14} for a discussion).
Here, $\lambda$ represents the geometric mean longitude, while $\varpi$ is the geometric 
longitude of pericentre. 
A ring of diffuse material discovered in $Cassini$ images \citep{Porc04,Porc05}, R/2004 S1, 
shares Atlas' orbit.
		
\citet{Spit06} fitted Voyager, HST and $Cassini$ ISS observations 
of Atlas spanning the period 2004 May to 2005 October. 
They found periodic perturbations in Atlas' orbit, 
with an amplitude of about 600 km along the orbital motion 
and a period of about 3 years, which they attributed to the 54:53 CER with Prometheus.  
They also identified the 70:67 mean motion resonance with Pandora
(with an amplitude $\sim 150$ km) and argued that since Prometheus and Pandora 
are interacting chaotically with each other, the orbit of Atlas itself might also be chaotic.
More recently, \citet{Coop15} extended the timespan of $Cassini$ ISS observations
(2004-2013) to fit the orbits and the masses of the 
inner satellites. This timespan extension allowed to cover the most recent 
chaotic interaction between Prometheus and Pandora in February 2013, and
the latest switch in the orbits of the co-orbitals Janus and Epimetheus in 
January 2010. They showed that Atlas 
is currently librating in both the 54:53 CER and the LER with Prometheus, 
making it yet another example of coupled CER/LER in the Saturn 
system, in common with Aegaeon, Anthe and Methone \citep{Elmout14,Spit06,Coop08,Hed09}, 
and possibly the Neptune's ring arcs \citep{Nam02,Ren14}. Using full numerical integrations 
combined with the FLI (Fast Lyapunov Indicator) time evolution, \citet{Coop15}  showed that the 
orbit of Atlas is chaotic, with a short Lyapunov time of about 10 years.

In this paper, we confirm that the origin of chaos for Atlas is the coupled resonant 
interaction (CER/LER) with Prometheus. The interactions between the two resonances is 
investigated using the CoraLin analytical model \citep{Elmout14}, showing that the chaotic 
zone fills almost all the corotation sites occupied by the satellite's orbit. 
We show that the four 70:67 overlapping apse-type mean motion resonances (see Table \ref{tab_reson}) 
due to Pandora have a much weaker effect on Atlas. We compare the results of the frequency analysis 
for the full numerical model fitted to $Cassini$ observations \citep{Coop15} 
and for a simplified system consisting of Saturn, Atlas and Prometheus only. The 
frequency analysis also allows us to study the effect of the coupling between the two 54:53 
resonances on the existence and relevance of proper frequencies. 

The paper is organized as follows. In Section \ref{sec:reson_summary}, we 
provide an overview of the most recent ephemerides and resonant perturbations 
of Atlas. Section \ref{sec:analytic} summarizes the analytical modelling based
the CoraLin model. This model is then numerically integrated in 
Section \ref{sec:numerical}, using the orbital parameters 
of Atlas. Section \ref{sec:freq_ana} focuses on the frequency analysis, 
and a summary and discussion are given in Section \ref{sec:discuss}.

\section{Resonant perturbations from the most recent ephemerides}
\label{sec:reson_summary}

\cite{Coop15} fitted a numerical model to new $Cassini$ ISS
astrometric data for Atlas, Prometheus, Pandora, Janus and Epimetheus. 
The state vectors and masses were solved at the epoch 2007 JUN 01 00:00:00.0, and 
ephemerides spanning the period 2000 to 2020 were generated. 
Using the parameters of Saturn given in Table \ref{tab_param_saturn}, 
the fitted state vectors are converted into geometric elements using the method 
of \citet{Ren06}, and the solutions are summarized in Table \ref{tab_orb_elements}. 
In Figure 1, we reproduce the longitude offsets of Atlas, Prometheus and Pandora
relative to a linear ephemeris using mean motion values of  
598.31312, 587.28501 and 572.78861 deg.day$^{-1}$, respectively. 
Sudden and anti-correlated changes in mean motion appear for Prometheus and 
Pandora. These jumps result from chaotic interactions between the two moons 
due to the overlap of four 121:118 apse-type mean motion resonances 
\citep{Gold03a,Ren03}. As concluded by \citet{Farm06}, the changes in mean motion 
do not always correlate with pericentre anti-alignments.  
The $\sim 1.7$ year oscillations of Pandora's mean longitude are due to 
the nearby 3:2 CER with Mimas. Pandora's semi-major axis lies approximately 50 km inside this
resonance, and also 180 km inside the 3:2 LER with Mimas \citep{French03}. 
Changes in mean motion are also apparent for Atlas, and the mean longitude is 
dominated by a $\sim 4$ year oscillation, as a result of the 54:53 CER with Prometheus. 

\cite{Coop15} showed numerically that the orbit of Atlas is chaotic with a Lyapunov time 
of order 10 years. On the other hand, since Prometheus and Pandora 
interact chaotically through 121:118 resonances, and since Atlas is perturbed by the 
54:53 CER/LER with Prometheus, then the mean motion ratio of Atlas and Pandora is
$n_{AT}/n_{PA}=1.044552$ (see Table \ref{tab_orb_elements}), which is close to 
the $70:67$ resonance. 
The figures 2 and 3 show the time variations of the resonance critical angles. 
Figure 2 displays the CER/LER arguments $\Psi_C$ and $\Psi_L$ on a timespan of 100 years. 
Between 2000 and 2020 both arguments are librating, except short episodes of circulation.
These episodes occur around 2006 (resp. 2013) for the LER 
(resp. CER), while simultaneously the CER (resp. LER) argument is librating. 
The four 70:67 apse-type resonance arguments due to 
Pandora are displayed in Figure 3 between 2000 and 2020. These critical
angles are : $ \Psi_1=70\lambda_{PA}-67\lambda_{AT}- 3\varpi_{PA} $,
$ \Psi_2=70\lambda_{PA}-67\lambda_{AT}- 2\varpi_{PA} -  \varpi_{AT} $,
$ \Psi_3=70\lambda_{PA}-67\lambda_{AT}-  \varpi_{PA} - 2\varpi_{AT} $ 
 and 
$ \Psi_4=70\lambda_{PA}-67\lambda_{AT}- 3\varpi_{AT} $. 
The four resonances overlap, but the separatrix
crossings (where the critical angles go from a circulation motion to libration, or from 
libration to circulation) are not clearly correlated with the times of 
pericentre anti-alignments between Prometheus and Pandora (vertical black 
dashed lines), or Atlas and Pandora (dotted). Furthermore, the effect of 
this third-order resonance with Pandora on the dynamics of Atlas 
is much weaker than the CER/LER with Prometheus, 
as shown in Table \ref{tab_reson} which lists the resonance libration rates and the 
perturbing function coefficients. 
From Figure \ref{Fig_resonATPA}, we note that the angle $\Psi_3$ is much closer to 
libration, on the timespan considered, than the three other critical arguments 
of the 70:67 resonance with Pandora.
The next section details the analytical modelling 
of the motion of Atlas perturbed by Prometheus, in the framework of the 
elliptic planar restricted three-body problem.

\section{Analytical Modelling}
\label{sec:analytic}

The mass ratio of Atlas and Prometheus is 0.036 \citep{Coop15}, and orbital 
inclinations for these two satellites are very small (Table \ref{tab_orb_elements}). 
Therefore the motion of Atlas can be well-approximated using the CoraLin model 
\citep{Elmout14}, which describes the behavior of a test particle near a horizontal 
first order mean motion resonance $m+1:m$ with a perturbing satellite 
($m$ integer, here $m=53$), in the frame of the elliptic, planar, restricted three-body problem. 
According to this model, the motion of Atlas is 
described by a two degree of freedom system involving the two resonance critical angles
$\Psi_C$ and $\Psi_L$, after averaging the equations of motion 
over the rapidly varying angles. Then the coupled effects of the two resonant 
terms (CER/LER) can be studied through the following Hamiltonian :

\begin{equation}
{\cal H} = \frac{1}{2}(J_{C} - 
J_{L})^{2} -  D J_{L} -  \varepsilon_{C} \cos(\Psi_C) -  
 \varepsilon_{L} h \rm{,}
\label{eq_hamil}
\end{equation}

with the equations of motion\footnote{Note that the time scale used in the model 
is $\tau = n_C t $, where $t$ is the usual time. Therefore, an object at CER has 
an orbital period $T=2\pi$, and the dots in the equations are the derivatives 
with respect to $\tau$.} : 

\begin{equation}
\left\{
\begin{array}{lll}
 \dot{J}_C= & \displaystyle{ - \partial {\cal H} / \partial \Psi_C } 
 = &  -\varepsilon_{C}\sin(\Psi_C)\\
 \dot{\Psi}_C= & \displaystyle{ + \partial {\cal H} / \partial J_C } = & J_C - J_L \equiv \chi \\
 \dot{h} = & \displaystyle{ + \partial {\cal H} / \partial k } = & -(J_C - J_L + D)  k  \\
  \dot{k} = & \displaystyle{ - \partial {\cal H} / \partial h } = & +(J_C - J_L+ D) h  + \varepsilon_{L}. \\
\end{array}
\right.
\label{eq_allchi}
\end{equation}

The various quantities entering in (\ref{eq_hamil}) and (\ref{eq_allchi}) are defined in 
Table~(\ref{tab_param_coralin}). The elements $a$, $e$ and $\dot{\varpi}$ denote 
the geometric semi-major axis, eccentricity and orbital precession rate 
(forced by the planet's oblateness), 
with subscripts $S$ for the perturbing satellite (here Prometheus), $a_C$ is the CER semi-major axis, 
$n_C$ is the corresponding mean motion, $\displaystyle \chi = \frac{3}{2} m \frac{a-a_C}{a} $ 
measures the Atlas' distance from the exact CER, and 
$M_S$ (resp. $M$) is the mass of the satellite (resp. the central body). The terms
$A^m$ and $E^m$ (cf.~Table \ref{tab_reson}) are combinations of Laplace coefficients \citep{Shu84}.
Here, these coefficients can be approximated  by $A^m  \sim -E^m \sim 0.8m$ since $|m|$ is large
(m=53). 
The first two equations of the system~(\ref{eq_allchi}) describe the CER and the 
last two ones the LER. The coupling between the two resonances arise from (i)
the $J_L$ term in the second equation, which states how the particle orbital eccentricity
driven by the LER perturbs the corotation pendulum motion, and (ii) the $J_C$ term in the 
third and fourth equations, which indicates how the CER affects the motion of the 
eccentricity vector (h, k) associated with the LER.  
The CoraLin model has three  fundamental parameters $D$, $\displaystyle \varepsilon_{C}$, 
$\displaystyle \varepsilon_{L}$ : 
$D$ is the (normalized) distance in frequency between the CER and the LER, 
indicative of the coupling, 
$\displaystyle \varepsilon_{C}$ is the CER strength
($\displaystyle n \sqrt{| \varepsilon_{C} |}$ the CER frequency), and 
$\displaystyle \varepsilon_{L}$ represents the LER eccentricity forcing.

%%%%%%%%%%%%%%%%%%%%%%%%%%%%%%%%%%%%%%%%%%%%%%%%%%%%%%%%%%%
%%%%%%%%%%%%%%%%%%%%%%%%%%%%%%%%%%%%%%%%%%%%%%%%%%%%%%%%%%%
%%%%%%%%%%%%%%%%%%%%%%%%%%%%%%%%%%%%%%%%%%%%%%%%%%%%%%%%%%%

For $D=0$, the CER and the LER are superimposed. Then 
the two degrees of freedom system (\ref{eq_allchi}) 
described by the Hamiltonian (\ref{eq_hamil})
admits a second integral of motion, and is thus integrable. 
This second integral was found by \citet{Ses84} for the general case 
of two non-zero masses, and extended to the restricted case by \citet{Wis86}, 
while being further analyzed by \citet{Hen86}. This result 
is rediscussed in \citet{Elmout14}. 
As $|D|$ increases, the coupling between the two resonances leads to 
chaotic motions as long as the LER radial location remains inside the 
CER site, see Figure 5 of \citet{Elmout14} and the next Section.  

For $|D|$ large, the resonances are decoupled and can be treated separately, 
see \citet{Elmout14}. 
Actually, neglecting the LER for $D$ large, the first two equations of the 
system~(\ref{eq_allchi}) reduce to:
\begin{equation}
\left\{
\begin{array}{ll}
\displaystyle
\dot{\chi} = -\varepsilon_{C}\sin(\Psi_C) \\
\displaystyle
\dot{\Psi_C}  = \chi.  \\
\end{array}
\right.
\label{eq_conservatif}
\end{equation}
This simple pendulum model describes, in other contexts, the libration of a satellite 
in a spin-orbit resonance \citep{Gold66}, or the Neptune's ring arcs confinement 
by the moon Galatea through a 42:43 resonance \citep{Gold86,Nam02}. 
Stable oscillations of $\Psi_C$ occur around $\Psi_C = 0$ (resp. $\Psi_C = \pi$) for $\varepsilon_{C}$ 
positive (resp. negative) with periods $2\pi / n_C$. The 
half-width of the CER site is given by 
$\displaystyle \frac{4}{3} a_C \frac{\sqrt{\varepsilon_{C}}}{|m|}$. 
Conversely, considering a perturbing satellite on a circular orbit, $\varepsilon_{C}=0$ and the CER 
vanishes. In this case $J_C$ is the Jacobi constant, and the system~(\ref{eq_allchi}) 
reduces to the classical second fundamental model for Lindblad resonance \citep{Hen83}:
\begin{equation}
\left\{
\begin{array}{ll}
\dot{h} = & -(J_C - J_L + D)  k  \\
\dot{k} = & +(J_C - J_L+ D) h  + \varepsilon_{L}. \\
\end{array}
\right.
\label{eq_lin}
\end{equation}

The CoraLin model can be easily modified to implement satellite orbital 
migrations and explore scenarios of capture into CERs. The 
analytical estimate of capture probabilities is not an easy task, 
in particular when the eccentricity of the perturbing satellite is large enough
so that the CER libration sites encompass the LER radius. 
\citet{Elmout14} discussed the case of the small Saturnian satellites 
Anthe, Methone and Aegaeon, captured into CERs with Mimas
($10:11$, $14:15$ and $7:6$, respectively).

\section{Numerical integrations}
\label{sec:numerical}

Here we present representative results of numerical integrations of the CoraLin model,  
in the case of Atlas perturbed by the $54:53$ CER/LER with Prometheus. 

Figure \ref{Fig_ae} shows a comparison of the orbital elements 
(semi-major axis and eccentricity) as a function of time for Atlas,  
derived both from CoraLin (in red) and
from a three-body simulation (black) including Atlas, Prometheus and 
Saturn's oblateness (up to $J_6$ included). The latter is actually 
presented in Figure 13 of \cite{Coop15}. 
The initial conditions are given in Table \ref{tab_elements_Fig_ae}, 
and are obtained from the ephemeris at epoch 
2000 JAN 01 12:00:00.0 UTC (JED 2451545.0) by converting state 
vectors to geometric elements using the algorithm of \cite{Ren06}.  
The integration of the 
averaged equations of motion of the Coralin model is in very good 
agreement with the full numerical model, confirming that the interactions 
of Atlas with Prometheus arising from the $54:53$ resonances grabs the 
essential parts of the dynamics.  

Surfaces of section ($\Psi_C$, $\chi$), showing the topology 
of the CoraLin system described by (\ref{eq_allchi}), are presented in Figure \ref{Fig_coralin}. 
In these sections, the positions of Atlas (in red) with respect to the CER radial location 
are plotted every time the $k$ component of the eccentricity vector 
is equal to zero. 
The reference radius $\chi=0$ corresponds to the CER radial location, and the  
LER radius at $\chi = -D$ is indicated in blue.
The surfaces of section start respectively on JED=2452647.6710 (2003 JAN 8, 04:06:14 UT), 
2454303.0976 (2007 JUL 21, 14:20:32 UT), 2456142.4397 (2012 AUG 2, 22:33:10 UT) and 
2458227.2788 (2018 APR 18, 18:41:28 UT), 
with initial orbital elements for Atlas and Prometheus derived from the three-body simulation
shown in Figure \ref{Fig_ae}. The values for Atlas' elements
are given in Table \ref{tab_elements_Fig_coralin}. 
To derive the location of the 54:53 CER ($a_C=137665.519$ km), 
we used the values for the mean motions and pericentre precession rates 
given in Table \ref{tab_orb_elements}, and computed iteratively the semi-major 
axis which cancels the derivative of the resonance argument. 
The satellite initial conditions, the CER radius $a_C$ and the mass ratio between Prometheus and 
Saturn (from Table \ref{tab_param_saturn} and \ref{tab_orb_elements}) are used 
to compute the values of the CoraLin parameters $D$, $\varepsilon_{C}$, 
$\varepsilon_{L}$ given in Table \ref{tab_param_coralin_atlas}. 
These parameters correspond to a CER half-width of 1.65 km (equivalent to $\Delta \chi= 1$), 
a distance between the CER and the LER locations $\displaystyle \frac{2}{3 |m|}a_C D \sim -0.36$ km, 
and a libration period $\displaystyle P_{\rm{lib}}=2\pi/(n \sqrt{|\varepsilon_{C}|})=3.45$ years.  
We note that the four satellites Anthe, Methone, Aegaeon and Atlas 
have very similar $\varepsilon_{L}$ values, see Table 2 of \citet{Elmout14}. 

The semi-major axis variations (Figure \ref{Fig_ae}) places Atlas in different parts 
of the CoraLin phase space. The surfaces of section (Figure \ref{Fig_coralin}) 
show that Atlas alternates the chaotic or regular motions, 
with semi-major axis variations of amplitude $\sim 1.5$ km comparable 
to the the CER half-width. This is a different
CoraLin regime compared to the cases of Aegaeon, Methone, Anthe which are embedded in arcs 
of material and have regular CoraLin orbits \citep{Elmout14}. 
From the twenty years simulation (Figure \ref{Fig_ae}), we estimate that episodes of chaotic motion 
(more precisely, orbital elements that correspond to chaotic orbits in the CoraLin phase space) 
add up to about 14 years. Note that we obtain the same phase portraits if we 
use initial conditions for Atlas and Prometheus derived from the ephemerides \citep{Coop15}, 
i.e. from integrations including the perturbations from all the other Saturnian satellites.

\section{Frequency analysis}
\label{sec:freq_ana}

A conservative dynamical system can be described by its frequencies \citep{Lask92}.
The frequency analysis is a method for studying the stability of orbits, 
based on a refined numerical search for a quasi-periodic approximation of 
its solutions over a finite time interval \citep{Lask90,Lask92,Lask93}.
For regular motions, this technique has the advantage of 
giving rise to an analytical representation of the solutions. 
It is also powerful for analysing weakly chaotic motion in 
hamiltonian systems.  In this case, the frequencies obtained are not well defined 
and thus vary in time, with a rate related to the chaos strength. 
On the other hand, determining the frequencies that 
have influence on the orbital elements of the Saturnian moons 
(and how those frequencies eventually change with time) is important, 
as this can be helpful to study in detail the moon interactions, the resonances
and their effects on the ring structures.  
In this aim we used the frequency analysis method, as described in e.g.  
\citet{LDV06}.

Despite the chaotic motion of Atlas, we can obtain relevant results 
on the frequencies of the system by selecting suitable time intervals. 
We compare here the results of the frequency analysis 
of the full numerical model fitted to $Cassini$ observations \citep{Coop15} 
and a simplified system consisting of Saturn, Atlas, Prometheus only. 
This allows us to confirm the results obtained both with the CoraLin model 
(Section \ref{sec:numerical}) or with the FLI simulations \citep{Coop15}.   

We have examined the following systems : 
\begin{itemize}
\item (1) the full numerical model fitted to $Cassini$ observations (Figure 1) 
\item (2) a 3-body simplified system with the same initial conditions but consisting 
of Saturn, Atlas, Prometheus only
\item (3) the same system as (2) but with an eccentricity for Prometheus 
$e_{PR}=2.8 \times 10^{-5}$
\item (4) the same system as (2) but with Prometheus' orbit circular. 
\end{itemize}

The dynamics of Atlas, perturbed by Prometheus, has the following 
characteristic timescales : (i) short periods ($\sim 14$ hours) associated with the 
orbital frequency, (ii) precession periods ($\sim 4$ months) associated with the 
apsidal precession rates $\dot{\varpi}$, 
(iii) libration periods ($\sim 4$ and $\sim 6$ years, respectively) associated 
with the CER and the LER, 
and (iv) a Lyapunov time of the same order ($\sim 10$ years). 

In case (4) (Prometheus' orbit circular), the CER is suppressed and in (3), 
the LER and the CER decouple since the Prometheus eccentricity value implies a
half-width for the CER site of 0.18 km, i.e. half the distance between the LER and the 
CER. Therefore, the motion of Atlas is regular in these two examples, with the 
main perturbation arising from the LER, and the frequency analysis is very efficient. 
For example, Table \ref{a_Atlas_case_4} gives the solution for the semi-major axis of 
Atlas in the case (4). 
Given the timescales of the problem, the frequency analysis is performed on a time interval of 
30 years, with a stepsize of 0.1 day.
The series arguments are easily identified using the three fundamental frequencies of motion 
(mean motions for Atlas and Prometheus $n_{AT}$ and $n_{PR}$, and LER libration frequency $\nu_L$). 
Thus, the series is quasi-periodic, 
meaning that the case (4) corresponds to a regular motion. Comparable solutions are achieved
for the other orbital elements of Atlas. 
For the case (3), where Atlas is trapped into the 54:53 LER and is close to but outside the CER, 
the analysis is less obvious because of the small eccentricity of Prometheus.
Nevertheless, we are able to clearly identify the four fundamental frequencies of the (two 
degrees-of-freedom) system : the three previous ones, $n_{AT}$, $n_{PR}$, $\nu_L$, 
and the Prometheus' pericentre precession rate ${\dot{\varpi}_{PR}}$.  
Furthermore, no significant variations of the frequencies (lower than 10$^{-5}$ deg.day$^{-1}$ 
for $n_{AT}$ or 10$^{-3}$ deg.day$^{-1}$ for $\nu_L$)
are found by shifting the time interval chosen for analysis (30 years) 
on a 200 year numerical integration. This confirms that the case (3) corresponds to a 
regular motion too.

As expected, the method fails for the more realistic cases (1) or (2), 
as the system is chaotic on short time scales, with a Lyapunov time comparable 
to the resonance libration periods. 
The frequencies $\dot{\Psi}_{C}$ and 
$\dot{\Psi}_{L}$ are separated by a small distance $\dot{\varpi}_S-\dot{\varpi} 
\simeq -0.124$ deg.day$^{-1}$, whereas the CER half-width corresponds 
to a frequency difference of $\sim 0.57$ deg.day$^{-1}$. This overlap leads to chaotic motion. 
Chaos is visible, for instance, in the random transitions of $\Psi_{C}$ from 
libration to circulation (or from circulation to libration). Such transitions lead to 
opposite separatrix crossings of $\Psi_{L}$, which block out the  
determination of the frequencies on a given time interval. This is illustrated in Figure 2
for the model fitted to $Cassini$ observations (case 1) and 
in Figure \ref{Fig_CER_LER_run3} for the 3-body system (case 2), 
both on a 100 year timespan. 
However, a partial representation of the frequencies for cases (1) and (2) can 
be obtained on time domains where no CER/LER separatrix crossings occur. The results
are summarized in Table \ref{frequencies_chaotic}. 
The method is much less efficient than in the regular case because of chaos and 
reduced time intervals. 
Nevertheless, mean motion values and pericentre precession rates are well determined, allowing 
us to verify the resonance conditions. 
We notice that the time intervals considered are of the same order as the Lyapunov time, 
as expected.
On the other hand, when the motion is sufficiently regular (i.e., no CER/LER transitions), 
the precision of the resonance rates (last column of Table \ref{frequencies_chaotic}) 
increases with the length of the time interval.  
The comparison of cases (1) and (2) shows that the essential part of Atlas' 
dynamics is controlled by the interactions with Prometheus due to the 54:53 resonances.

We can identify the third-order resonant perturbations due to Pandora. 
Frequency analysis for the case (1) on the timespan 2006-2020 
(where the 54:53 LER argument is librating, see Table \ref{frequencies_chaotic}) 
leads to $\dot{\varpi}_{PA}=2.599742 $ deg.day$^{-1}$ and a variation for 
$ \Psi_3=70\lambda_{PA}-67\lambda_{AT}-  \varpi_{PA} - 2\varpi_{AT} $ of $0.0430$ deg.day$^{-1}$. 
This value is about ten times smaller than the variations of the three other 70:67 arguments, 
confirming that the angle $\Psi_3$ seen in Figure \ref{Fig_resonATPA} looks closer to libration. 
Furthermore, separatrix crossings for $\Psi_3$ occur at the same epochs as those of the 
54:53 LER argument $54 \lambda_{PR} - 53 \lambda_{AT} - \varpi_{AT}$. Further work is needed
to explain if this is purely coincidental or not. 

We have also verified that by increasing the eccentricity of Prometheus,  
e.g. by a factor of ten ($e_{PR}=2 \times 10^{-2}$), Atlas' motion becomes extremely chaotic 
with a very short Lyapunov time, and an impossibility to compute the main frequencies. 
This was already shown in \citet{Elmout14} when the resonances are superimposed ($D<<1$). 
 
When the eccentricity of the perturbing satellite is small enough so that the two resonances
are well separated, i.e. when the CER half-width is smaller than the 
distance of the LER from the CER ($\varepsilon_{C}<D$), the motion becomes regular. Here, the 
limit case between regular and chaotic motions corresponds to a small eccentricity 
$e_{PR} \sim 8 \times 10^{-5}$. 

\section{Summary and Discussion}
\label{sec:discuss}

Using initial states and masses fitted to new $Cassini$ ISS observations, \citet{Coop15}
developed an improved high-precision numerical model for the orbits of Atlas, Prometheus, 
Pandora, Janus and Epimetheus. Based on this model, we confirmed that the orbit of Atlas is 
chaotic, as a consequence of the coupled interaction between the 54:53 CER and LER with Prometheus. 
We showed that the chaotic region fills almost all the CER site occupied by Atlas' orbit, and
highlighted the 70:67 overlapping resonances with Pandora.  
The frequency analysis allowed us to confirm our results, showing that the dynamics of Atlas
is mostly 
controlled by the 54:53 resonant perturbations from Prometheus. A partial representation 
of the frequencies of motion was obtained on timespans comparable to the Lyapunov time of the 
system, where no separatrix crossings occur. We showed that Atlas motion is chaotic as soon 
as Prometheus' eccentricity exceeds a value of about $8 \times 10^{-5}$. 
The smallness of this value suggests that the CER/LER coupling and the resulting chaotic motions 
could be a relatively common process during the orbital evolution of small, nearby moons 
around Saturn, as the satellite orbits expand through the transfer of angular momentum 
from the rings and cross numerous mean motion resonances. 
Similar chaotic interactions could be frequent as part of the orbital evolution 
of satellites around other giant planets, and require further 
investigation. Indeed, recent works showed that the closely-packed Uranian system of 
inner low-mass satellites is configured in chains of interlinked first- and second-order 
eccentric resonances, contributing to chaotic motions \citep{French12,Quillen14,French15}.  
It also remains to assess the possible consequences of the dynamical results presented here 
on the orbital evolution timescales.

\acknowledgments

This work was supported by the Science and Technology Facilities Council (Grant No. ST/M001202/1) 
and Cooper and Murray are grateful to them for financial assistance. 
Murray is grateful to The Leverhulme Trust for the award of a Research Fellowship.
Cooper thanks the University of Lille 1 for additional funding while he was an invited 
researcher at the Lille Observatory.
Cooper, El Moutamid, Murray, Renner and Vienne thank the Encelade working group for interesting
discussions.

\clearpage

\begin{deluxetable}{lll}
\tablecaption{Saturn constants, from \citet{Coop15}.}
\tablewidth{0pt}
\tablehead{Constant&Value&units}
\startdata
 $GM$  & $3.793120706585872 \times 10^{7}$  &km$^3$ s$^{-2}$\\
 Radius          &60330  &       km\\
 $J_{2}$         &$1.629084747205768 \times 10^{-2}$ & \\
 $J_{4}$         &$-9.336977208718450 \times 10^{-4}$& \\
 $J_{6}$         &$9.643662444877887\times 10^{-5}$ & \\
\enddata
\label{tab_param_saturn}
\end{deluxetable}

\begin{deluxetable}{lrrr}
\tablecaption{Geometric orbital elements for Atlas, Prometheus and 
Pandora at epoch 2007 JUN 01 00:00:00.0 UTC (JED 2454252.50075446), 
computed from fits to $Cassini$ observations \citep{Coop15}. 
The elements $a$, $e$, $i$, $\Omega$, $\varpi$, $\lambda$ are 
respectively the 
semi-major axis, the eccentricity, the inclination, the longitude of ascending node, 
the longitude of pericentre, and the mean longitude. 
The mean motion and the pericentre precession rate are computed 
self-consistently from the semi-major axis using the Saturn constants
given in Table \ref{tab_param_saturn}. 
}
\tablewidth{0pt}
\tablehead{ & Atlas & Prometheus & Pandora }
\startdata
Mass (kg) & $5.751 \times 10^{15}$ & $1.600 \times 10^{17}$ & $1.368 \times 10^{17}$  \\
\hline
$a$ (km) & 137664.946 & 139378.239 & 141711.251  \\
\hline
$n$ (deg.day$^{-1}$) & 598.316026   &  587.283454  & 572.796769   \\
\hline
$e$  & 0.00114  &  0.00222  & 0.00417   \\
\hline
$i$ (deg) & 0.00290  &  0.00753 & 0.05024   \\
\hline
$\Omega$ (deg) & 21.19790   &  86.59026   & 339.90039   \\
\hline
$\varpi$ (deg) & 325.55527  &  263.32452 &  52.27079      \\
\hline
$\dot{\varpi}$ (deg.day$^{-1}$) & 2.881135   & 2.757159     & 2.599218     \\
\hline
$\lambda$ (deg) & 310.40476   &  50.69084    & 281.02045     \\
\hline
\enddata
\label{tab_orb_elements}
\end{deluxetable}

\begin{deluxetable}{lrrl}
\rotate 
\tablecaption{Resonance arguments, rates, periods, coefficients. The rates
are the time derivatives $\dot{\Psi}$ of the resonance angles, computed using the mean 
motion values and the pericentre precession rates given in Table \ref{tab_orb_elements}, and 
the periods are $2\pi/\dot{\Psi}$. The corresponding terms of the disturbing 
potential are given in the last column. The coefficients $f_i$ are combinations of 
Laplace coefficients \citep{Shu84}, $A^m=-\big{[} f_{27} + e^2_{AT} f_{28} + e^2_{PR} f_{29} \big{]}$, 
$E^m=-\big{[} f_{31} + e^2_{AT} f_{32} + e^2_{PR} f_{33} \big{]}$, keeping the terms up to order 
3 in eccentricities in the potential and evaluating the coefficients $f_i$ at 
$\alpha=a_{AT}/a_{PR}$ in the tables of \citet{md99}.}
\tablewidth{0pt}
\tablehead{  Argument & Rate (deg.day$^{-1}$) & Period (yr) & Coefficient
($\times 10^{-9}$ m$^2$.s$^{-2}$) }
\startdata
$ \Psi_L=54\lambda_{PR}-53\lambda_{AT}-\varpi_{AT} $ & -0.324494 & 3.03743 
& $\displaystyle (G m_{PR}/{a_{PR}}) e_{AT} A^m = 3.7976 $   \\
$ \Psi_C=54\lambda_{PR}-53\lambda_{AT}-\varpi_{PR} $ &  -0.200518 & 4.91540 
& $\displaystyle (G m_{PR}/{a_{PR}}) e_{PR} E^m = - 7.4585 $  \\ 
\hline
$ \Psi_1=70\lambda_{PA}-67\lambda_{AT}- 3\varpi_{PA} $ &  0.800002  & 1.23203 
& $\displaystyle (G m_{PA}/{a_{PA}}) e^3_{PA} f_{85} = 0.0847 $ \\ 
$ \Psi_2=70\lambda_{PA}-67\lambda_{AT}- 2\varpi_{PA} -  \varpi_{AT}   $  & 0.518085 & 1.90244 
& $\displaystyle (G m_{PA}/{a_{PA}}) e^2_{PA} e_{AT} f_{84} = -0.0680 $ \\ 
$ \Psi_3=70\lambda_{PA}-67\lambda_{AT}-  \varpi_{PA} - 2\varpi_{AT}   $  & 0.236169  & 4.17339 
& $\displaystyle (G m_{PA}/{a_{PA}}) e_{PA} e^2_{AT}  f_{83} =  0.0182 $ \\ 
$ \Psi_4=70\lambda_{PA}-67\lambda_{AT}- 3\varpi_{AT} $ & -0.0457478 & 21.5448 
& $\displaystyle (G m_{PA}/{a_{PA}}) e^3_{AT} f_{82} = - 0.0016 $ \\ 
\hline
\enddata
\label{tab_reson}
\end{deluxetable}

\begin{deluxetable}{|l|c|}
\tablecaption{Variables and parameters used in the CoraLin model
\citep{Elmout14}, see text Section \ref{sec:analytic} 
for details. The coefficients $A^m$ et $E^m$ are defined in the caption 
of Table \ref{tab_reson}.}
\tablewidth{0pt}
\tablehead{Quantities&Definitions}
\startdata
$ h $ & $ \sqrt{3} \mid m\mid e \cos(\Psi_L) = \sqrt{2J_{L}} \cos(\Psi_L) $  \\
\hline
$ k $ &  $ \sqrt{3} \mid m\mid e \sin(\Psi_L)= \sqrt{2J_{L}} \sin(\Psi_L) $ \\
\hline
$ J_L $ & $ (h^2+k^2)/2 = 3m^2e^2/2$  \\
\hline
$ J_C $ & $\chi + J_L = 3 m (a-a_C)/(2a_C) + J_L$  \\
\hline
$ \varepsilon_{L} $ & $  \sqrt{3} \mid m\mid (M_S/M) (a_C/a_S) A^m$  \\
\hline
$ \varepsilon_{C} $ & $ 3m^2 (M_S/M) (a_C/a_S)  E^m e_S $  \\
\hline
$ D $ & $ (\dot{\varpi}_S - \dot{\varpi})/n_C $  \\
\enddata
\label{tab_param_coralin}
\end{deluxetable}

\begin{deluxetable}{lrr}
\tablecaption{Initial conditions for the simulations of Figure \ref{Fig_ae}.
}
\tablewidth{0pt}
\tablehead{ & Atlas & Prometheus }
\startdata
$a$ (km) & 137666.519 & 139378.180 \\
\hline
$e$  & 0.00117  &  0.00222   \\
\hline
$\varpi$ (deg) &  87.97515  &  357.21193      \\
\hline
$\lambda$ (deg) & 116.02563   &  203.51184        \\
\hline
\enddata
\label{tab_elements_Fig_ae}
\end{deluxetable}

\begin{deluxetable}{lr|r|r|r}
\tablecaption{Initial conditions for Atlas used for the surfaces of section of 
Figure \ref{Fig_coralin}. Prometheus moves on an unperturbed orbit with elements
given in Table \ref{tab_orb_elements}.    
}
\tablewidth{0pt}
\tablehead{ & 2003 JAN 8  & 2007 JUL 21  & 2012 AUG 2  & 2018 APR 18     
}
\startdata
$a$ (km)& 137664.290    
& 137665.545
& 137666.696 
& 137664.800 \\
\hline
$e$  & 0.00118 
& 0.00105  
& 0.00106  
& 0.00109   \\
\hline
$\varpi$ (deg) &  259.75152    
&  345.53665  
&  243.35908  
&  119.44170     \\
\hline
$\lambda$ (deg) & 111.26215    
& 216.09571   
& 199.10884  
& 181.89609           \\
\hline
\enddata
\label{tab_elements_Fig_coralin}
\end{deluxetable}

\begin{deluxetable}{lll}
\tablecaption{CoraLin parameters for Atlas. For comparison with
Anthe, Methone and Aegaeon, the values are also
provided in the nomenclature of \citet{Elmout14}, where $\varepsilon_{C} \equiv 1$. }
\tablewidth{0pt}
\tablehead{$\varepsilon_{C}$ & $D$ & $\varepsilon_{L}$}
\startdata
  $-2.28 \times 10^{-7}$ & $-2.07 \times 10^{-4}$ & $1.11 \times 10^{-6} $  \\
 1  & -0.43 & 0.11 \\
\enddata
\label{tab_param_coralin_atlas}
\end{deluxetable}

\begin{deluxetable}{lrrrc}
\tablecaption{Frequency analysis for the semi-major axis of Atlas in the case 
(4) of Section \ref{sec:freq_ana} (Prometheus on a circular orbit). 
The time is from 2007 JUN 01 00:00:00.0 UTC. 
The time interval used is 30 years with a stepsize of 0.1 day.
The three fundamental frequencies are used for the identification of the arguments of the series terms. 
These frequencies are the LER libration frequency $\nu_L$, the mean mean motion 
of Atlas $n_{AT}$, and the one of Prometheus $n_{PR}$. 
The frequency values derived from this run are respectively $\nu_L=0.153114$ deg.day$^{-1}$, 
$n_{AT}=598.309930$ deg.day$^{-1}$ and $n_{PR}=587.283438$ deg.day$^{-1}$.
The series is expressed in cosine.}
\tablewidth{0pt}
\tablehead{Number & Amplitude & Frequency        & Phase  & Identification   \\
        &        (km) & (deg.day$^{-1}$) & (deg)  &      }
\startdata
  1           &  137665.87876 &     0              &      0       &  -               \\
  2           &           0.87952  &     0.153114 &  179.83 &   $                  \nu_L$ \\ 
  3           &           0.03816  &     0.459343 &  179.49 &   $                 3 \nu_L$ \\ 
  4           &           0.01224  &    10.873434 & -100.51 &   $    n_{AT} -     n_{PR}  -  \nu_L$ \\ 
  5           &           0.01188  &    11.179667 &    79.11 &   $    n_{AT} -     n_{PR} +  \nu_L$ \\
  6           &           0.01094  &     0.306233 &   179.65 &  $                         2   \nu_L$ \\ 
  7           &           0.01054  &    22.052890 &   160.11 &  $2  n_{AT} - 2  n_{PR}             $ \\ 
  8           &           0.00946  &    33.079410 &    59.64 &   $3  n_{AT} - 3  n_{PR}             $ \\ 
  9           &           0.00868  &    44.105919 &   -40.77 &   $4  n_{AT} - 4  n_{PR}             $ \\  
 10          &           0.00806  &     55.132426 & -141.13 &   $5  n_{AT} - 5  n_{PR}             $ \\ 
... \\
\enddata
\label{a_Atlas_case_4}
\end{deluxetable}

\begin{deluxetable}{lrrrrcr}
\tablecaption{Frequency analysis for the chaotic cases (1) and (2) of Section \ref{sec:freq_ana}.
The values are in deg.day$^{-1}$.  
The last column gives the rates $54  n_{PR} - 53  n_{AT}-\dot{\varpi}_{AT}$ (LER) 
or $54  n_{PR} - 53  n_{AT}-\dot{\varpi}_{PR}$  (CER), i.e. the
difference with the exact resonance condition. }
\tablewidth{0pt}
\tablehead{Time interval &   $ n_{AT}$   &    $n_{PR}$        &    $\dot{\varpi}_{AT}$   
                         &  $\dot{\varpi}_{PR}$  & LER/CER & Rate }
\startdata
Case (1) \\
  2000-2012         &  598.314000 &     587.284665         &      2.879580       &  2.757720            & CER  &     -0.0278       \\
  1950-2010			&  598.311138 &     587.282364		   &      2.881033       &  2.757703            & CER  &     -0.0004       \\
  2006-2020         &  598.311003 &     587.285110         &      2.879606       &  2.757686            & LER  &      0.0332       \\
  Case (2) \\
  2050-2080           &  598.312204 &     587.283466         &      2.879486       &  2.757168            & CER  &      0.0031       \\
  2020-2080           &  598.312277 &     587.283466         &      2.879626       &  2.757168            & CER  &     -0.0007       \\
  2000-2016           &  598.309703 &     587.283467         &      2.878903       &  2.757168            & LER  &      0.0140       \\
  2087-2102           &  598.309497 &     587.283465         &      2.880308       &  2.757168            & LER  &      0.0235 
  \enddata
\label{frequencies_chaotic}
\end{deluxetable}

\clearpage
\setlength{\topmargin}{0in}
\setcounter{figure}{0}

%%%%%%%%%%%%%%%%%%%%%%%%%%%%%%%%%%%%%%%%%%%%%%%%%%%%%%%%%%%%%%%%%%%%%
\begin{figure}
\epsscale{0.6}
\plotone{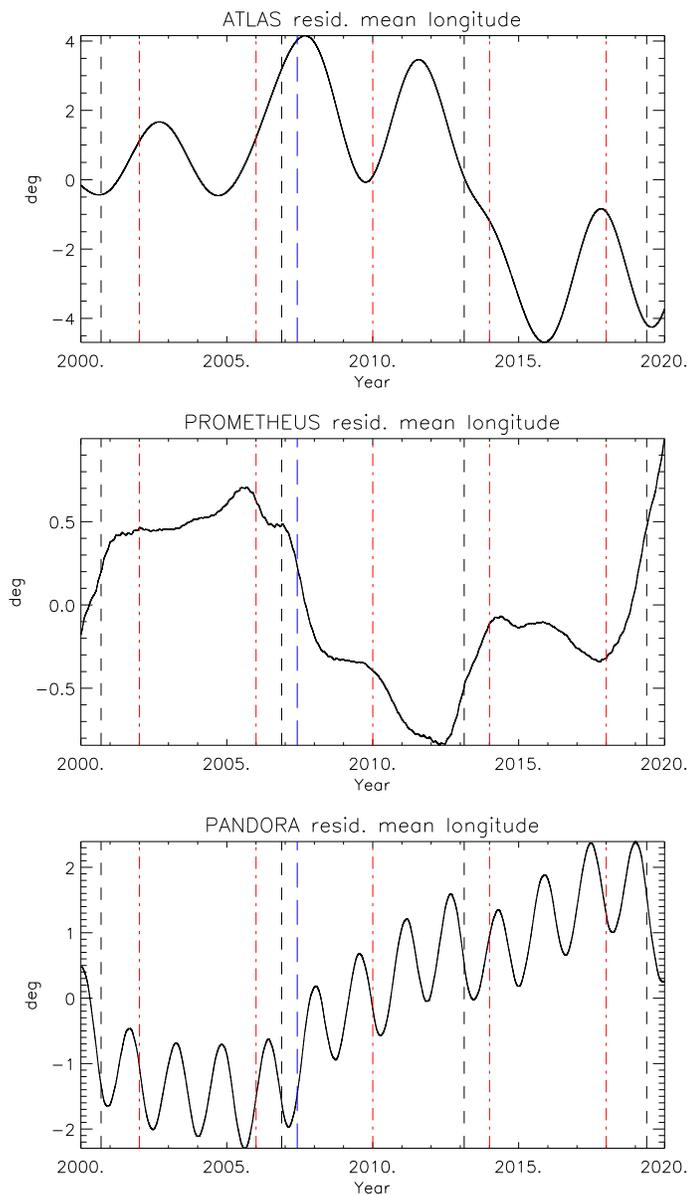}
\caption{Residual Mean Longitudes for (a) Atlas, (b) Prometheus and (c) Pandora between 
2000 and 2020 from the full numerical model of \citet{Coop15}. 
$Cassini$ astrometric observations spanning 2004 February to 2013 August have been used to 
fit the orbits and the satellite masses. 
Linear background trends have been subtracted from the mean longitudes using 
rates of (a) 598.31312 deg/day, (b) 587.28501 deg/day and (c) 572.78861 deg/day. 
Vertical black dashed lines mark times of closest approach between 
Prometheus and Pandora. 
Vertical red dot-dashed lines mark times of switches in the configuration of 
Janus and Epimetheus. 
Vertical blue line marks fit epoch.
This figure is a reproduction of Figure 10 from \citet{Coop15}, provided
here for convenience.}
 \label{Fig_lambda}
\end{figure}

%%%%%%%%%%%%%%%%%%%%%%%%%%%%%%%%%%%%%%%%%%%%%%%%%%%%%%%%%%%%%%%%%%%%%
\begin{figure}
\centering
  \includegraphics[width=.7\textwidth]{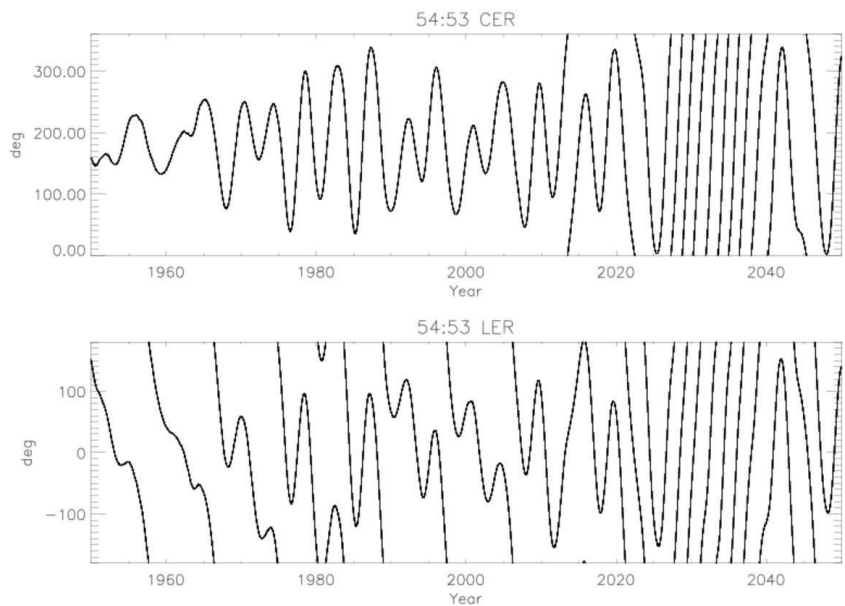}
\caption{$54:53$ CER and LER arguments for Atlas and Prometheus  
from the full numerical model of \citet{Coop15}, extended to 
100 years between 1950 and 2050. The angles are 
(a) $ \Psi_C=54\lambda_{PR}-53\lambda_{AT}-\varpi_{PR} $ and
(b) $ \Psi_L=54\lambda_{PR}-53\lambda_{AT}-\varpi_{AT} $. 
Random transitions of $\Psi_C$ from libration to circulation (or from circulation to 
libration) occur, with opposite separatrix crossings of $\Psi_L$, 
e.g. around 2013 or 2022. 
}
\label{Fig_resonATPR}
\end{figure}

%%%%%%%%%%%%%%%%%%%%%%%%%%%%%%%%%%%%%%%%%%%%%%%%%%%%%%%%%%%%%%%%%%%%%
\begin{figure}
\centering
  \includegraphics[width=.75\textwidth,angle=90]{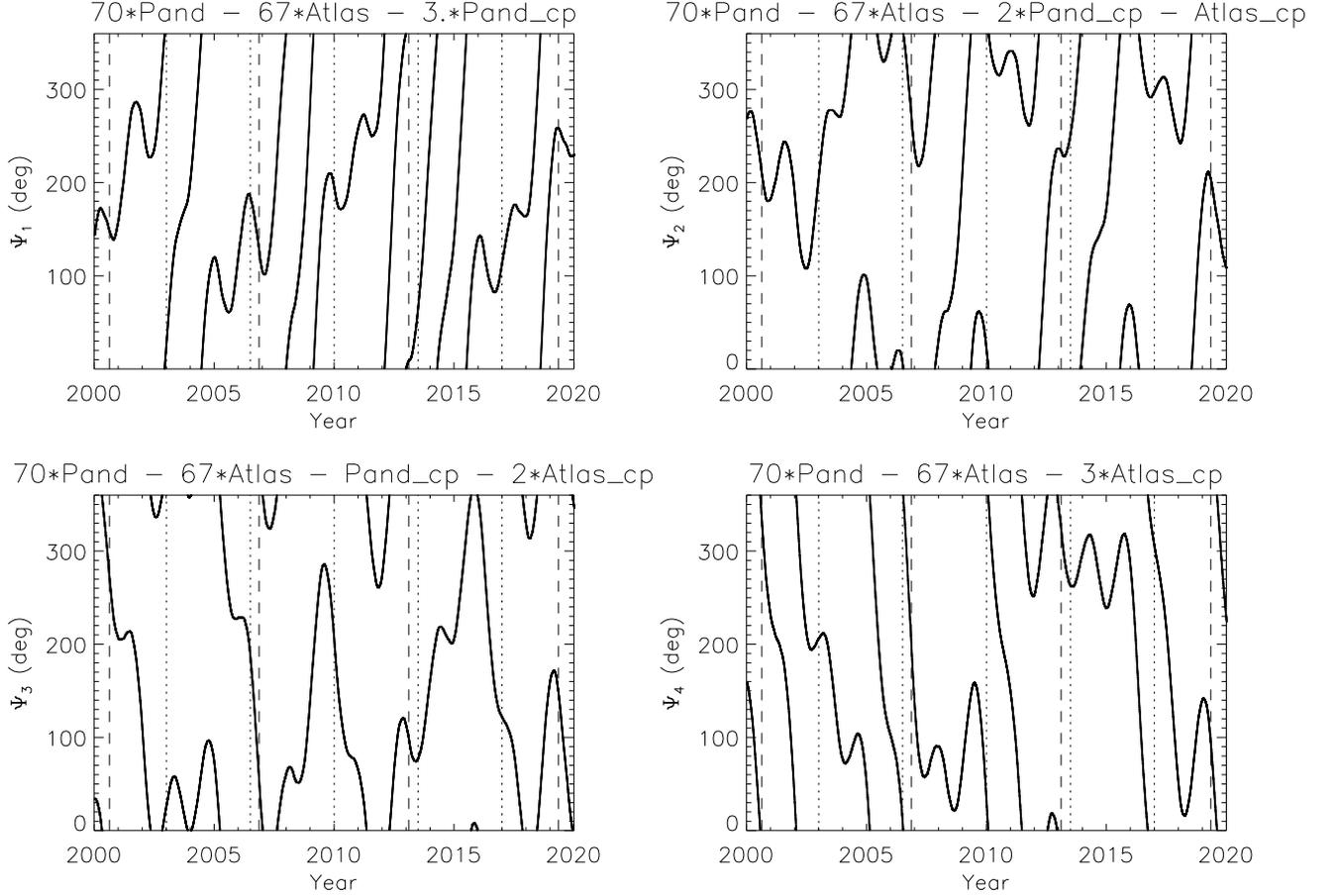}
\caption{$70:67$ resonance arguments for Atlas and Pandora between 2000 and
2020 from the full numerical model of \citet{Coop15}. 
The angles are 
(a) $ \Psi_1=70\lambda_{PA}-67\lambda_{AT}- 3\varpi_{PA} $,
(b) $ \Psi_2=70\lambda_{PA}-67\lambda_{AT}- 2\varpi_{PA} -  \varpi_{AT} $, 
(c) $ \Psi_3=70\lambda_{PA}-67\lambda_{AT}-  \varpi_{PA} - 2\varpi_{AT} $ and 
(d) $ \Psi_4=70\lambda_{PA}-67\lambda_{AT}- 3\varpi_{AT} $. 
Vertical black dashed (resp. dotted) lines mark times of apse anti-alignments 
between Prometheus and Pandora (resp. Atlas and Pandora). 
The four resonances overlap, but the separatrix crossings are not clearly correlated 
with the times of pericentre anti-alignments. The angle $\Psi_3$ is much
closer to libration than the three other resonance arguments.  
}
  \label{Fig_resonATPA}
\end{figure}

%%%%%%%%%%%%%%%%%%%%%%%%%%%%%%%%%%%%%%%%%%%%%%%%%%%%%%%%%%%%%%%%%%%%%
\begin{figure}
\centering
  \includegraphics[width=.5\textwidth,angle=90]{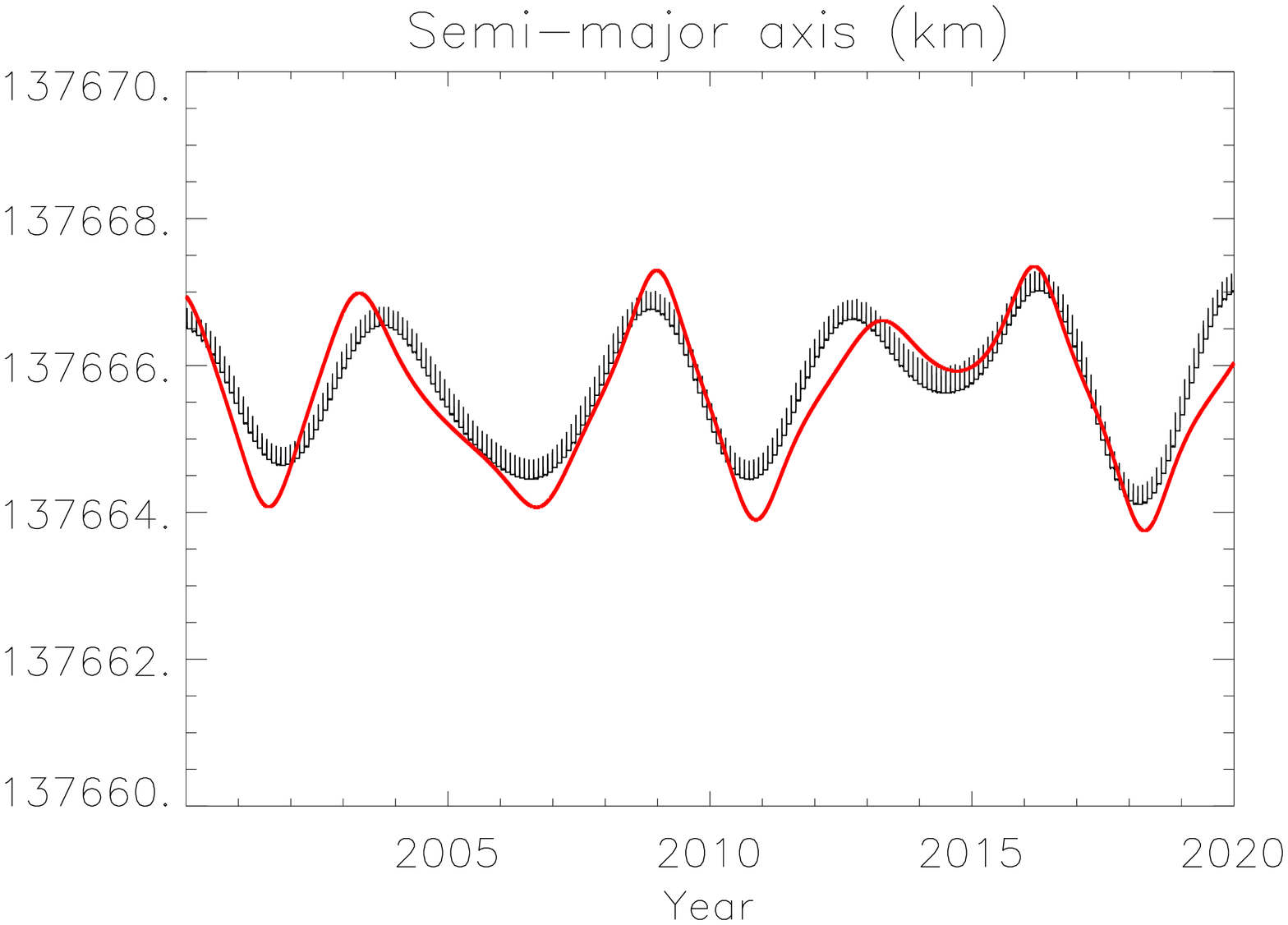}
  \includegraphics[width=.5\textwidth,angle=90]{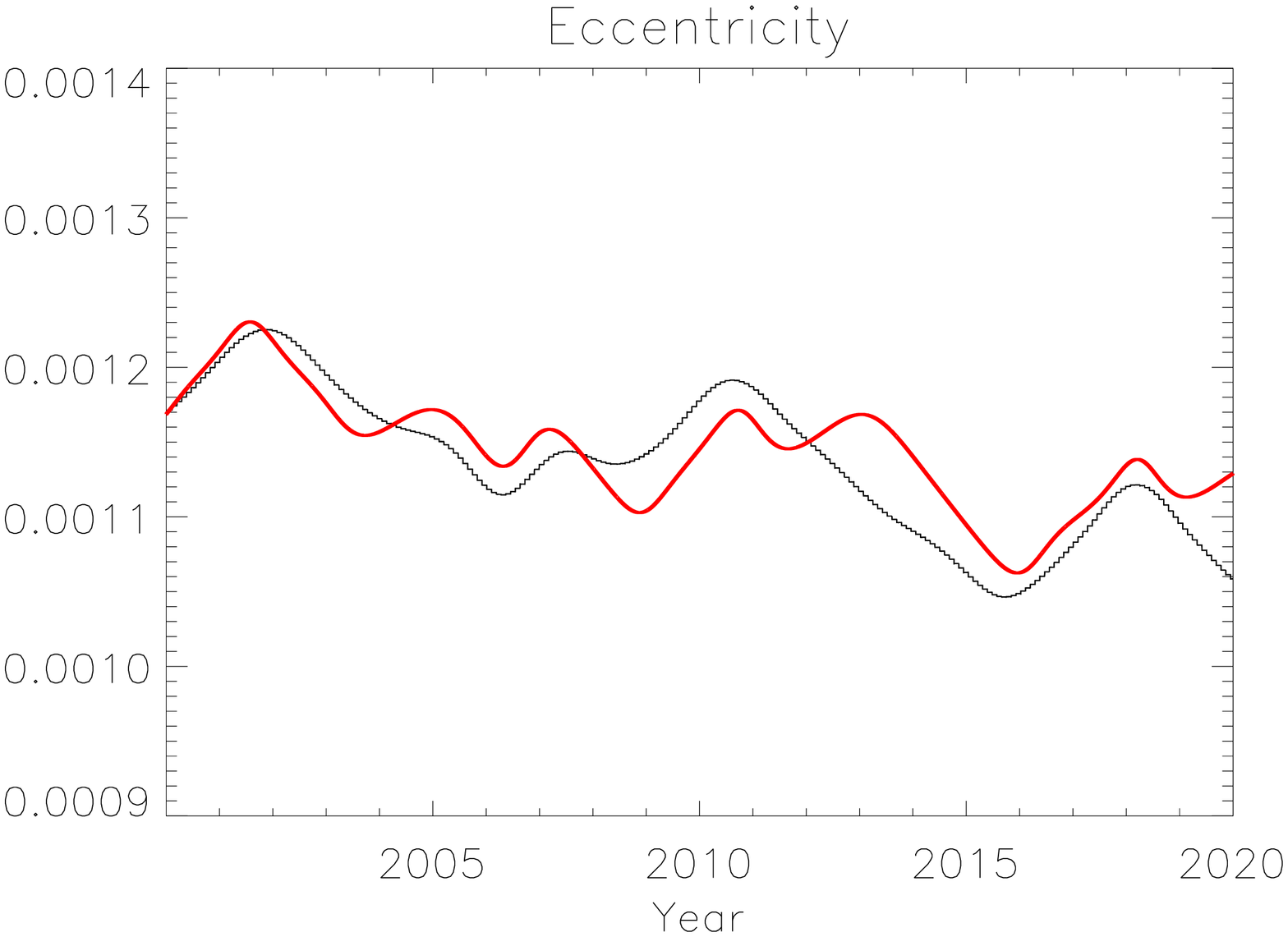}
\caption{Semi-major axis and eccentricity for Atlas between 2000 and 2020. 
The red curve is given by the CoraLin model, and the black curve is 
from a full numerical integration including Prometheus and Saturn's oblateness up to and 
including terms in $J_6$.}
\label{Fig_ae}
\end{figure}

%%%%%%%%%%%%%%%%%%%%%%%%%%%%%%%%%%%%%%%%%%%%%%%%%%%%%%%%%%%%%%%%%%%%%
\begin{figure}
\centering
  \includegraphics[width=.35\textwidth]{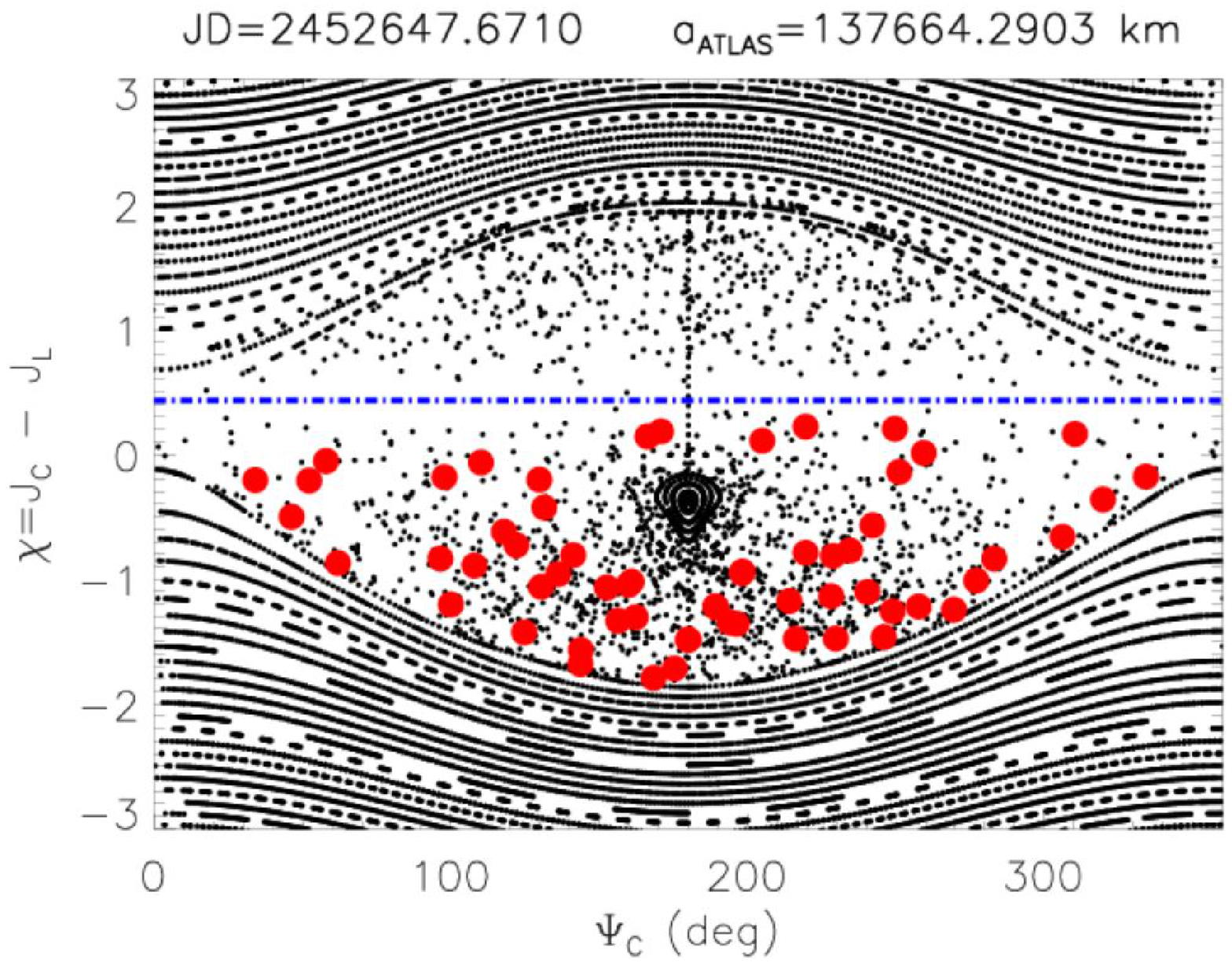}
  \includegraphics[width=.35\textwidth]{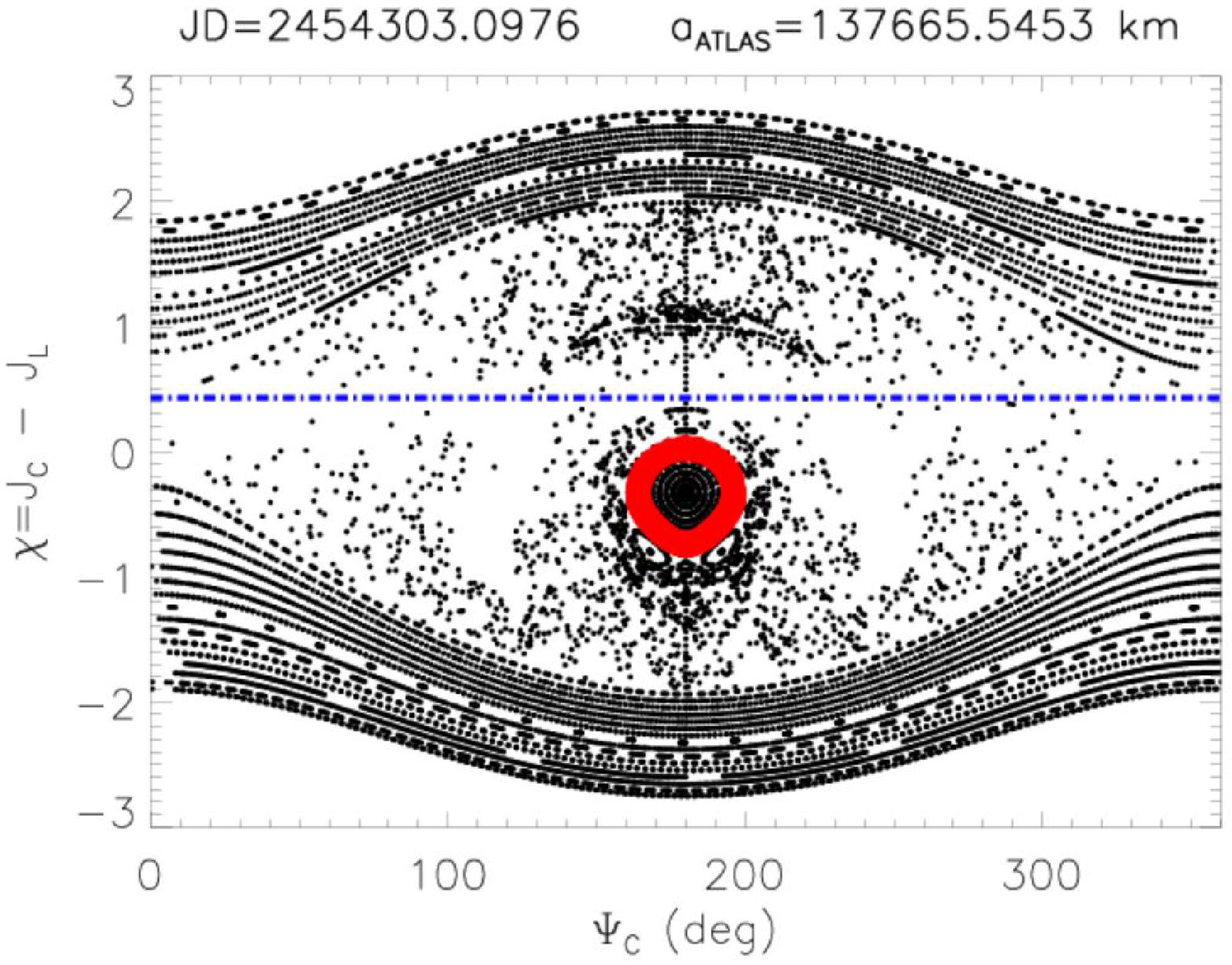} \\
  \includegraphics[width=.35\textwidth]{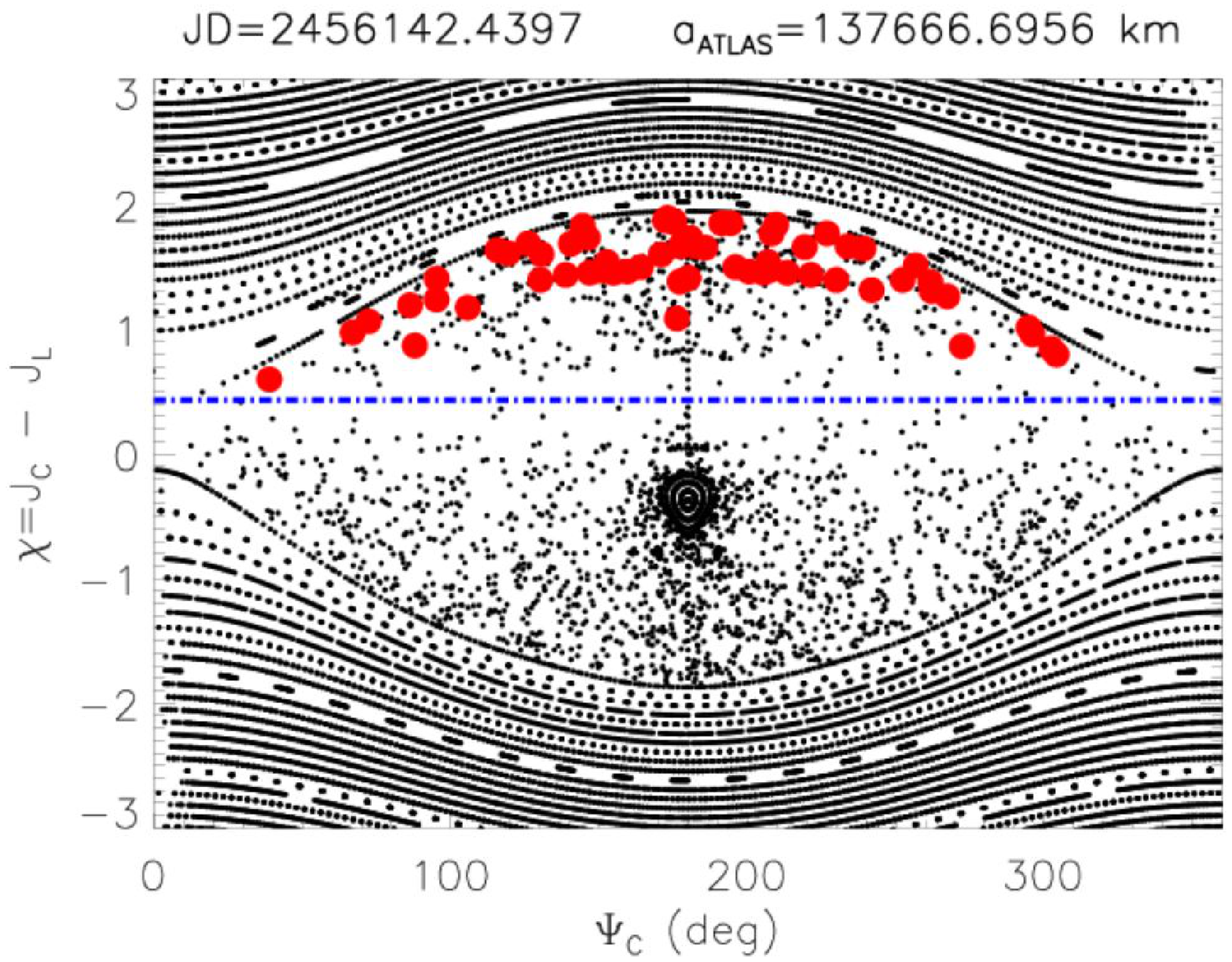}
  \includegraphics[width=.35\textwidth]{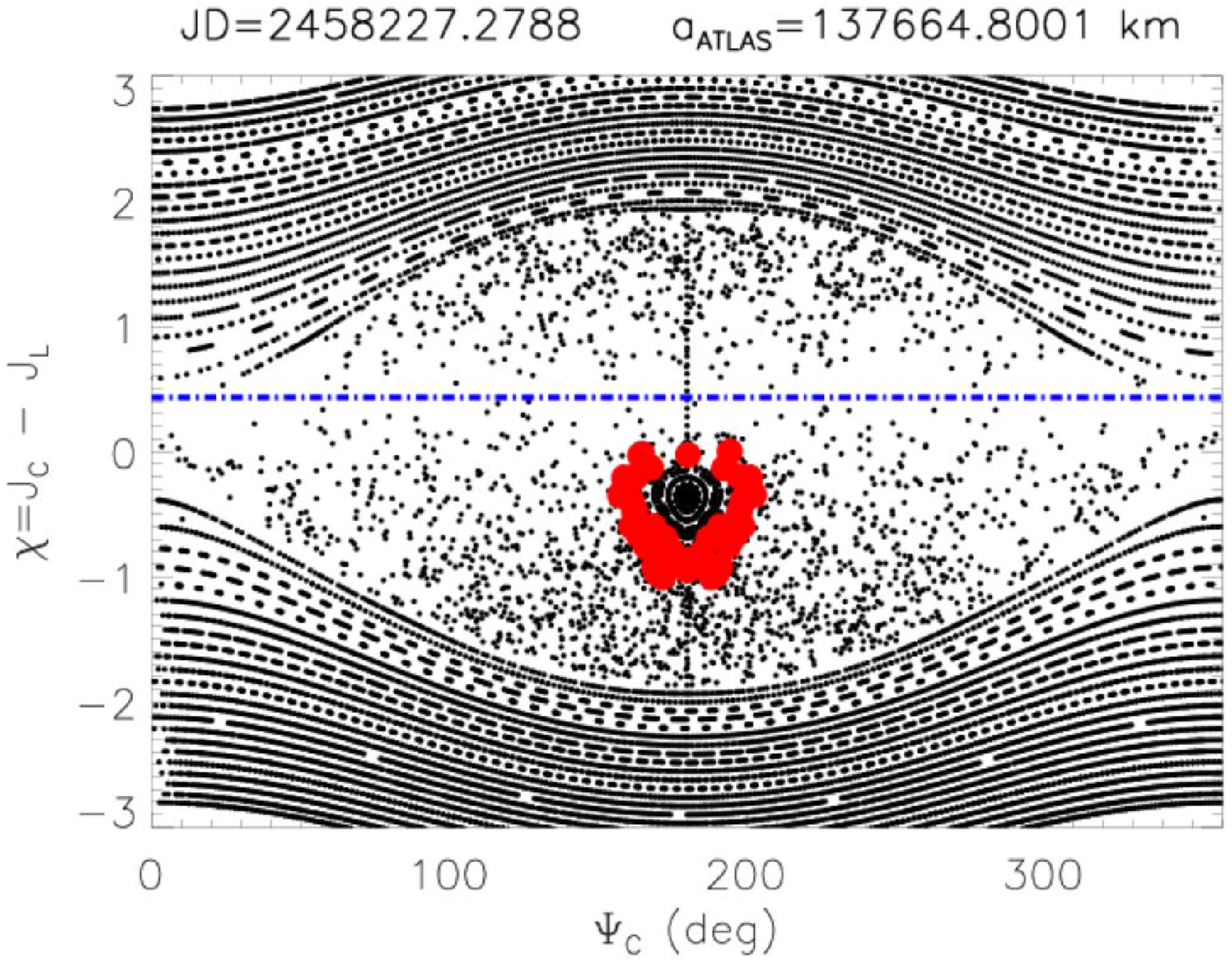}
\caption{CoraLin surfaces of section ($\Psi_C$, $\chi$) 
starting on 2003 JAN 8, 2007 JUL 21, 2012 AUG 2, 2018 APR 18. The LER radius at $\chi = -D$ is in blue.
Atlas is in red, using initial conditions (see Table \ref{tab_elements_Fig_coralin}) 
from the from the three-body simulation Saturn, Prometheus, 
Atlas shown in Figure \ref{Fig_ae}. For each surface, the integration time is 550 years.}. 
\label{Fig_coralin}
\end{figure}

%%%%%%%%%%%%%%%%%%%%%%%%%%%%%%%%%%%%%%%%%%%%%%%%%%%%%%%%%%%%%%%%%%%%%
\begin{figure}
\centering
  \includegraphics[width=.7\textwidth,angle=90]{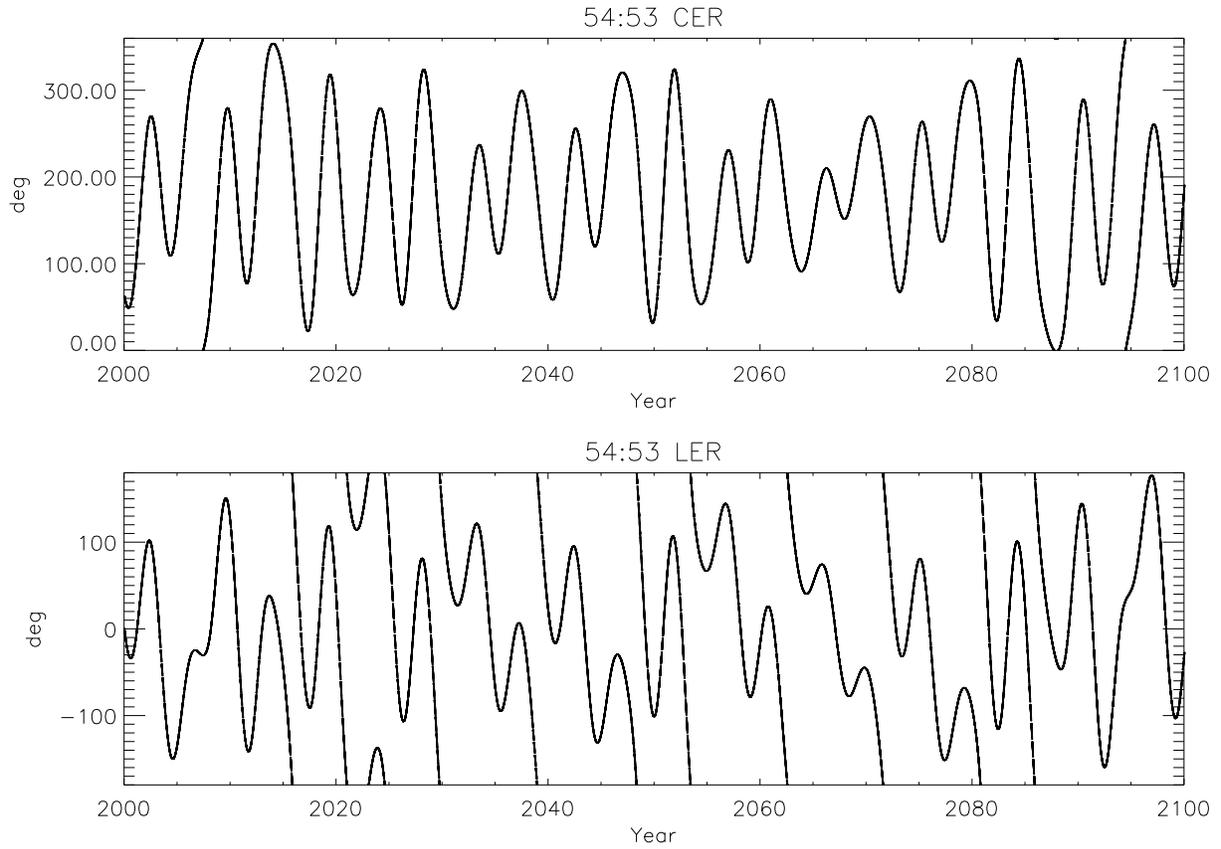}
\caption{$54:53$ CER and LER arguments for Atlas and Prometheus between 2000 and 2100 
from a three-body model consisting of Atlas, Prometheus and Saturn 
(case 2 of Section \ref{sec:freq_ana}).  
As in Figure \ref{Fig_resonATPR}, chaos is visible in the random CER/LER separatrix 
crossings, which prevent the determination of frequencies on arbitrary time intervals,
see Table \ref{frequencies_chaotic}. 
}
  \label{Fig_CER_LER_run3}
\end{figure}

\end{document}